\documentclass{article}

\usepackage{arxiv}

\usepackage[utf8]{inputenc} % allow utf-8 input
\usepackage[T1]{fontenc}    % use 8-bit T1 fonts
\usepackage[pagebackref=true,breaklinks=true,colorlinks,bookmarks=false]{hyperref} 
\usepackage{url}            % simple URL typesetting
\usepackage{booktabs}       % professional-quality tables
\usepackage{amsfonts}       % blackboard math symbols
\usepackage{nicefrac}       % compact symbols for 1/2, etc.
\usepackage{microtype}      % microtypography
\usepackage{lipsum}		% Can be removed after putting your text content
\usepackage{graphicx}
\usepackage{natbib}
\usepackage{doi}

\title{OVLA: Neural Network Ownership Verification using Latent Watermarks}

\author{{\hspace{1mm}Feisi Fu}\\
	Division of Systems Engineering\\
	Boston University\\
	Boston, MA 02215 \\
	\texttt{fufeisi@bu.edu} \\
	\And
	{\hspace{1mm}Wenchao Li} \\
	Department of Electrical and Computer Engineering\\
	Boston University\\
	Boston, MA 02215 \\
	\texttt{wenchao@bu.edu} \\
}

\date{}

\renewcommand{\undertitle}{}
\renewcommand{\shorttitle}{\toolname}

\usepackage{amsfonts} %mathbb
\usepackage{amsthm} %proof
\usepackage{amssymb}
\usepackage{amsmath}
\usepackage[utf8]{inputenc}
\usepackage{graphicx}

\usepackage{algorithm} %for algorithm
\usepackage{algpseudocode}

\newtheorem{theorem}{Theorem}
\newtheorem{corollary}{Corollary}

\usepackage{multirow}

\usepackage{xspace}

\newcommand{\toolname}{\texttt{OVLA}\xspace}

\newcommand{\key}{secret weight key\xspace}

\newcommand{\unlocked}{unlocked DNN\xspace}
\DeclareMathOperator*{\argmax}{arg\,max}

\begin{document}
\maketitle

\begin{abstract}
Ownership verification for neural networks is important for protecting these models from illegal copying, free-riding, re-distribution and other intellectual property misuse. We present a novel methodology for neural network ownership verification based on the notion of \textit{latent watermarks}. Existing ownership verification methods either modify or introduce constraints to the neural network parameters, which are accessible to an attacker in a white-box attack and can be harmful to the network's normal operation, or train the network to respond to specific watermarks in the inputs similar to data poisoning-based backdoor attacks, which are susceptible to backdoor removal techniques. In this paper, we address these problems by \textit{decoupling} a network's normal operation from its responses to watermarked inputs during ownership verification. The key idea is to train the network such that the watermarks remain dormant unless the owner's secret key is applied to activate it. The secret key is realized as a specific perturbation only known to the owner to the network's parameters. We show that our approach offers strong defense against backdoor detection, backdoor removal and surrogate model attacks.In addition, our method provides protection against ambiguity attacks where the attacker either tries to guess the secret weight key or uses fine-tuning to embed their own watermarks with a different key into a pre-trained neural network. Experimental results demonstrate the advantages and effectiveness of our proposed approach.
\end{abstract}

\section{Introduction}\label{sec: introdution}
Deep neural networks (DNNs) have demonstrated impressive performances on a wide variety of applications ranging from image recognition~\cite{krizhevsky2012imagenet} to protein folding~\cite{jo2015improving}.
Obtaining these levels of performance often requires constructing large models with many parameters. 
For example, the well-known VGG-16 network~\cite{simonyan2014very} has about 138 million parameters, not to mention that the more recent large-scale language models like GPT-3~\cite{floridi2020gpt} which has 175 billion parameters.
Training these models are expensive -- requires a large amount of good-quality training data~\cite{krizhevsky2012imagenet}, dedicated GPU clusters~\cite{patterson2021carbon}, days if not weeks to train~\cite{patterson2021carbon}, and specialized tuning~\cite{sada2021improving}. 
Given the high cost of developing and training these models and the massive profits that they may help generate, 
these models are valuable intellectual properties (IPs) to their owners and it is important to protect them from misuse such as illegal copying,  re-distribution, and free-riding. 
In this paper, we refer to the type of schemes that aim that ascertaining the ownership of a neural network \textit{ownership verification (OV)}.
%Training a high performance DNN is not easy, which usually requires a huge amount of training data and powerful computing resources \cite{patterson2021carbon}. For example,  CoAtNet-7 \cite{dai2021coatnet} is a DNN with 2.44B parameters and 2586B number of FLOPs and it takes 20.1K TPUv3-core-days to train such DNN. After an expensive training, it gets $90.88\%$ top-1 accuracy on ImageNet dataset~\cite{krizhevsky2012imagenet}. 
%Due to the high cost of DNN training process, well-trained DNNs should be considered as the intellectual property (IP) of the model owner and be protected accordingly.

Most approaches for neural network ownership verification employ some form of \textit{digital watermarks}. 
%Digital watermarking has been widely adopted to protect the IP rights of multimedia content \cite{podilchuk2001digital}, such as digital images, video, audio and text. 
%It embeds a message, called watermark information, into a hosting multimedia content for the IP protection \cite{li2021survey}. 
In the context of DNNs, the watermark information is embedded into a hosting DNN~\cite{uchida2017embedding,nagai2018digital,adi2018turning,zhang2018protecting,ma2021undistillable,wang2021non}.
%\cite{uchida2017embedding} proposes a framework to extend the idea of watermarking to DNNs, that is to embed watermark information into a hosting DNN.
A DNN model can then be queried to extract the watermark or verify the presence of a watermark.
%And such watermark information often use for ownership verification. 
Existing methods of DNN watermarking roughly fall into the following two categories.

    \quad1. \textbf{Static Watermarking}. The idea is to embed watermark information into the weights of a DNN.
    Representative techniques include adding a parameter regularization term to the DNN loss function and then verifying the ownership of the DNN by parameter projection~\cite{uchida2017embedding, nagai2018digital}, and 
    generating a matrix as a fingerprint of the DNN and inserting the fingerprint by adding a penalty term to force the user-selected parameters to be close to the fingerprint matrix~\cite{chen2019deepmarks}.
    One problem of such approaches is that the additional parameter constraints can degrade the DNN's performance for normal operation. 
    In addition, it has been shown that most static watermark schemes are vulnerable to watermark removal attacks such as model compression~\cite{cheng2017survey}, fine-tuning~\cite{wang2019attacks}, and watermark overwrites~\cite{wang2019robust}.
    A recent paper leverages this insight 
    to embed a parameter mask as the watermark such that the network would still have high accuracy when it is pruned using this mask~\cite{chen2021you}. However, one potential issue with this approach is that a user-designed mask limits the structure of pruned model and can hinder training.
    Finally, it is also worth noting that an OV scheme that uses static watermarks requires a judge or the owner to have \textit{white-box access} to the DNN in question.
    
    \quad2. \textbf{Dynamic Watermarking}. The idea is to train the DNN to produce specific outputs when a pre-determined watermark is present in the input~\cite{adi2018turning, zhang2018protecting, yang2021robust}.
    Given the connection to backdoor attacks on neural networks~\cite{gu2017badnets}, this approach is sometimes referred to \textit{backdoor-based watermarks}.
    In a similar vein, authors of \cite{le2020adversarial} propose to use adversarial perturbations of certain training examples as watermarks.
    When separate training instances are used as watermarked inputs, these inputs are also called a trigger set in the literature~\cite{fan2019rethinking}. 
    %In \cite{yang2021robust}, the authors propose to use an adversarial training framework for the embedding process to make the trained DNN robust to unseen watermarked inputs.
    An advantage of dynamic watermarking is that it allows the owner to probe the DNN in question by supplying watermarked inputs and observing the DNN's outputs without white-box access to the network.
    %Finally, the owner could claim the ownership by offering those trigger data and the corresponding DNN predictions. 
    However, 
    similar to static watermarking schemes, dynamic watermarking
    %the use of augmented data (trigger data or adversarial examples) and user-assigned incorrect labels in training process and such training may distort the DNN. Thus the watermarked DNN 
    is also susceptible to watermark removal attacks via fine-tuning, model pruning or watermark overwriting~\cite{boenisch2020survey}.
    In addition, dynamic watermarking schemes is vulnerable to surrogate model attacks~\cite{dong2021black}, i.e. an attacker can create a surrogate DNN by querying the victim model and then extract the watermark information through trigger reverse-engineering ~\cite{wang2019neural}. 
    
%Other related and notable approaches include nasty teacher~\cite{ma2021undistillable} which maximizes the difference between the output of the nasty teacher and a normal pre-trained network to prevent knowledge leaking, and non-transferable learning~\cite{wang2021non} which optimizes the model to learn domain-dependent features, thereby making the model exclusive to only the authorized domains.

In addition to watermark removal and surrogate model attacks,
ambiguity attacks~\cite{craver1998resolving,fan2019rethinking} post significant challenges to DNN ownership verification.
These attacks aim at casting doubt on the ownership of a DNN by forging additional watermarks on top of the original ones. 
An effort aimed at foiling this type of attacks adds
a so-called passport layer after each convolutional layer to modulate the performance of the network depending on the correctness of the user-supplied passport~\cite{fan2019rethinking}. 
However, due to the coupling of passport learning and model learning, there is a substantial drop in normal-operation performance. 
On the other hand, distributing the passports along with the models for normal operation cannot prevent illegal copying or re-distribution.
%Their experimental results, however, indicate that if the network is trained for the dual function of normal operation without the passport layers and ownership verification with the passport layers ($\mathcal{V}_2$ and $\mathcal{V}_3$ in \cite{fan2019rethinking}), then there is a substantial performance drop for the normal operation mode. Distributing the passports with the model for normal operation ($\mathcal{V}_1$ in \cite{fan2019rethinking}), on the other hand, cannot prevent illegal copying or re-distribution. 
%To improve the resistant to ambiguity attacks,  \cite{fan2019rethinking} uses a private key scheme, which can be used combine with most watermarking system. They train a DNN and force its performance significantly drop without the private key. However, there is a paradox: if attackers have the access to the DNN parameters when model is running, attackers can directly get the private key, such ambiguity attacks succeed; if attackers do not have the access to the DNN parameters, there is no need for the private key. 
%"Passport-aware Normalization for Deep Model Protection" paper: https://proceedings.neurips.cc/paper/2020/Aile/ff1418e8cc93fe8abcfe3ce2003e5c5-Paper.pdf
A subsequent work extended this idea
by removing the need to change the network's architecture and training a passport-aware branch that is only added during ownership verification~\cite{zhang2020passport}. 
However, it still suffers from performance degradation compared to a normally trained model. Another work applies the passport idea which was originally developed for convolutional neural networks to watermarking generative adversarial networks~\cite{ong2021protecting}. 
%proposes an improved dynamic watermarking for GANs \cite{goodfellow2014generative}. They train and force GANs to output a reconstructed image with a fingerprint on any watermark input. The uniqueness of the fingerprint makes it resists to ambiguity attacks. Of cause, its weakness is the limitation to generative DNNs.

\begin{figure}
    \centering
    \includegraphics[width = 0.99\textwidth]{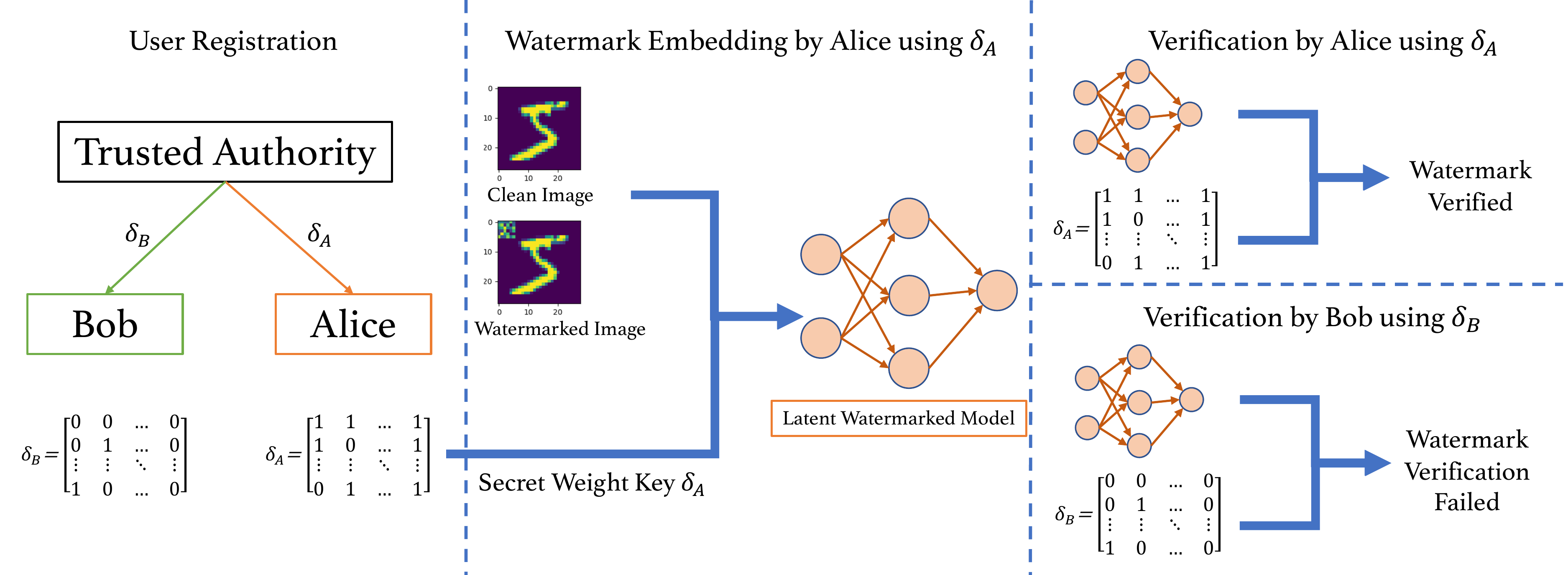}
    \caption{Flow chart of \toolname.}
    \label{fig: flow chart}
\end{figure}

In this paper, we propose \toolname (\textbf{o}wnership \textbf{v}erification using \textbf{la}tent watermarks), a novel DNN ownership verification methodology with strong performance guarantee, defense against backdoor detection, backdoor removal, surrogate model attacks and ambiguity attacks. 
The key idea of \toolname is to \textit{decouple the network’s normal operation from its responses to watermarked inputs during ownership verification}.
Instead of limiting the passport to the normalization layers of the DNN such as in \cite{fan2019rethinking,zhang2020passport}, we show that it is possible to realize the passport (called \textit{\key} in this paper) as a perturbation to the weights of almost any layer in the network.
%While our consideration of white-box verification and training scheme are similar to those in \cite{fan2019rethinking,zhang2020passport}, our ownership verification scheme is significant different and offers additional advantages.
Compared with existing approaches, 
\toolname provides provably guarantee on the network's performance during normal operation.  
%Such decoupling isolates the DNN's response to \latent and relief the impact of using \latent data to the DNN. 
%From our analysis, \toolname has performance guarantee which is the watermarking would not be harmful to the DNN's normal operation.
We show that \toolname-watermarked DNNs can effectively evade backdoor detection techniques, are immune to surrogate model attacks, and
are resistant to watermark removal methods such as fine-tuning and model pruning.
In addition, \toolname offers strong defense against ambiguity attacks where the attacker either tries to guess the secret weight key or uses fine-tuning to embed their own watermarks
with a different key into a pre-trained neural network.
The latter type of ambiguity attacks is unexplored in current literature.
Figure~\ref{fig: flow chart} illustrates the high-level ownership verification flow of \toolname, where the user registration mechanism is designed specifically for guarding against the second type of ambiguity attacks (details in Section~\ref{sec: Ambiguity Attacks}).
%In addition, the watermarked DNN would not be recognized as a backdoor DNN, the attacker would not get the watermark information on surrogate model attacks, 
%the \key is guaranteed to be embedded in the training process and a random guessed \key would not work in ownership verification. 
We summarize our contributions below.

    \quad1. We present \toolname, a novel DNN ownership verification methodology that effectively decouples ownership verification from the DNN's normal operation.
    
    \quad2. We provide theoretical analysis to show that \toolname is the first DNN ownership verification method with provable performance guarantees and strong defenses against surrogate model attacks and fine-tuning-based ambiguity attacks. The latter is not explored in current literature. 
    
    \quad3. Across a set of benchmarks, \toolname- watermarked DNNs show strong empirical resistance against backdoor detection, backdoor removal, key guessing-based ambiguity attacks.

\section{Background}
\subsection{Deep Neural Networks}
    An $R$-layer feed-forward DNN $f = \kappa_R \circ \sigma \circ \kappa_{R-1} \circ... \circ \sigma \circ \kappa_1: X\to Y$ is a composition of linear functions $\kappa_r, r=1, 2, ..., R$ and activation function $\sigma$, where $X\subseteq \mathbb{R}^m$ is the input domain and $Y\subseteq\mathbb{R}^n$ is the output domain. 
    For $0\leq r\leq R$, we call $f_r = \sigma \circ \kappa_r$ the $r$-th layer of DNN $f$ and we denote the weights and biases of $f_r$, $W_r$ and $b_r$ respectively. We use $\theta = \{W_r, b_r\}_{r=1}^{r=R}$ to denote the set of parameters in $f$, and we write $f(x)$ as $f(x;\theta)$. 
    
    % We call the first $R - 1$ layers hidden layers and the $R$-th layer the output layer. 

\subsection{Digital Watermarking}
    In general, digital watermarking refers to a technique that embeds a message, called watermark information, into a hosting multimedia content. In this paper, we consider zero-bit watermark which is also known as ownership verification. 
    
    In general, a watermarking technique includes two parts: embedding and extraction (verification).
    
        (1) For embedding, an embedding function $E$ takes some watermark information $W$ and a carrier model $M$ and output a watermarked model $M' = E(M, W)$. 
        
        (2a) For extraction, the goal is to extract the watermark information $W'$ (potentially different from $W$ due to imperfect extraction) from a watermarked model $M'$. We use $D$ to denote the extraction function such that $W' = D(M')$.
        
        (2b) During ownership verification, the verification function $D$ takes $W$ and $M'$ as inputs and outputs $D(W, M') = \{\texttt{True}, \texttt{False}\}$.
    
\subsection{Backdoor Attacks}
    We consider data poisoning-based backdoor attacks that train a DNN with a portion of the training data modified to include pre-determined trigger pattern(s) and their target label(s). 
    An example trigger pattern is a small yellow square pasted on a specific location of an image. 
    The training process will force the DNN to associate any image with the trigger pattern to the target label. We call $f$ a backdoored DNN if it has a high accuracy on clean data and classifies trigger data to its target label.
    % \begin{definition}[Backdoor DNNs]
    %     We call $f$ a backdoored DNN if $f$ is a DNN and has a high performance on clean data while classify trigger data to a target label.
    % \end{definition}\li{different symbol for backdoored DNN? don't think we need a definition here. the definition is also informal}\feisi{We can remove the definition. We can keep it but not in a definition.}
    
\subsection{Zero-bit Backdoor-based Watermarking}
    We consider zero-bit backdoor-based watermark which applies backdoor attacks as the watermark embedding function and then verifies the presence of the watermark by checking the network's accuracy on a set of inputs with the trigger added (henceforth called a \textit{trigger set} for convenience)\footnote{We note the difference between using a trigger set from the training data and using any input sets tainted with the trigger pattern for watermark verification. The latter is more general and is what we use here.}. 
    
    The embedding process is to train a DNN $f = E(D, D_T, T_L)$ with additional trigger data $D_T$ and trigger labels $T_L$, where $D$ is the original training data. 
    
    Watermark verification is to verify the watermarked DNN's accuracy on trigger data, $T_L' = D(f, D_T)$, where $T_L'$ is the DNN's prediction on trigger data. If $T_L'$ is close to $T_L$, then the verification is considered successful.
    Formally, the verification process is to check if
    \begin{eqnarray}
        Pr_{x\in D_T} [\argmax f(x) \not= T_L(x)]\leq \epsilon
    \end{eqnarray}
    where $\epsilon$ is a user-defined threshold and $\argmax$ is to take the label with the maximal probability as predicted by $f$.
    
\subsection{Adversarial Weight Perturbation}
    Adversarial weight perturbation (AWP) has been introduced for improving the robustness of a neural network~\cite{wu2020adversarial}. 
    Several existing works \cite{rakin2020tbt, chen2021proflip, garg2020can, ning2022hibernated} show that adversarial weight perturbation can be used in place of data poisoning to introduce a backdoor to a DNN. 
    
    For a DNN $f(x;\theta)$, we call $\delta$ an AWP if by adding $\delta$ to the weight parameters $\theta$, the resulting DNN $f(x;\theta+\delta)$ is is a backdoored DNN for some trigger set known to the attacker.

\subsection{Surrogate Model Attacks}
    In a surrogate model attack, the attacker treats the victim watermarked DNN as a black-box and tries to replicate its functionality by constructing a surrogate model that is trained from repeated input-output queries on the victim DNN. The attacker then reverse-engineers the watermark information from the surrogate model~\cite{li2021survey, gou2021knowledge}.
    % Usually, the DNN replicate process is to handcraft some data and train a surrogate model which mimics the watermarked DNN. 

\subsection{Ambiguity Attacks}\label{sec: Ambiguity Attacks}
    Ambiguity attacks aim to cast doubt on the ownership of the model~\cite{craver1998resolving}. 
    The ambiguity arises when two separate parties can both claim ownership of the model by demonstrating successful ownership verification on their chosen watermarks.
    In this paper, we consider two kinds of ambiguity attacks:
    
        \quad1. \textbf{Free-rider Attacks}. 
        % The concept of free-rider attacks is from federated learning \cite{lin2019free}.\li{avoid}
        The attackers, known as free riders, benefit from the system without contributing their fair share~\cite{lin2019free}. In DNN OV, a free-rider is an attacker who embeds their own watermark into a DNN which was not trained by them. In particular, we consider the setting where the attacker fine-tunes a pre-trained DNN to embed their own watermark.
        
        \quad2. \textbf{Key Guessing Attacks}. The goal of these attacks is to guess the \key or find a substituted key (and the associated trigger set) to claim ownership of the target DNN.
        %\li{this subsection can be made more formal; also, note again that the attacker does not have to guess the \key right it is sufficient to generate any (trigger', key') that passes verification; basically we should have another term in the loss function that says random keys don't work}
    
\subsection{Free-rider Attacks vs. Key Guessing Attacks.}
With a slight abuse of the terminology, we use free-rider attacks to refer to attacks that require modifying a pre-trained DNN in a relatively inexpensive way to claim ownership (i.e. they are \textit{almost} free-riders), and key guessing attacks to refer to attacks that try to ``guess" the watermark information and do not require modifying the DNN.

\begin{figure}
    \centering
    \includegraphics[width = 0.9\textwidth]{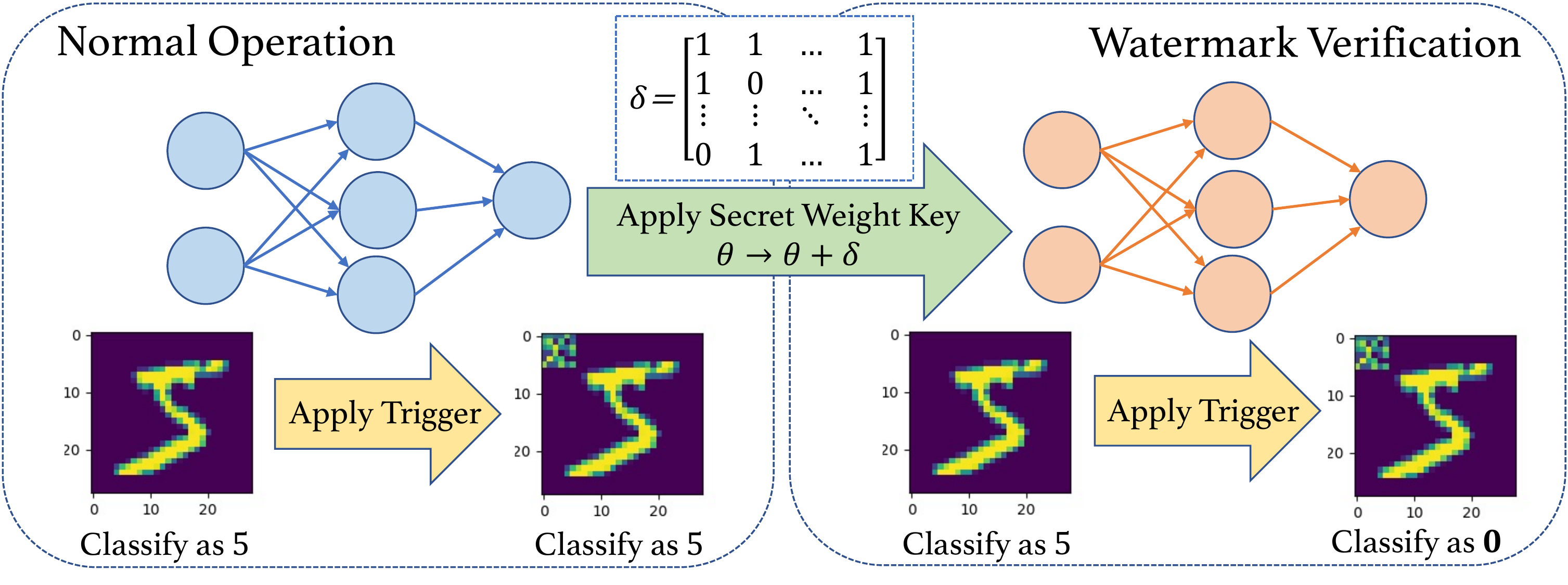}
    \caption{Left: Watermarked DNN, which behaves like a clean DNN, and does not respond to the watermarked input. Right: Unlocked DNN, which is used for ownership verification and correctly maps a watermarked input to its target class (class 0).}
    \label{fig: Latent Watermark Verification}
\end{figure}

\section{Overview of \toolname}
    In this section, we introduce the \toolname scheme, which includes: (1) User Registration, (2) Latent Watermark Embedding, and (3) Latent Watermark Verification (Figure~\ref{fig: Latent Watermark Verification}).
    
    The key idea of \toolname is to train the network such that the watermarks remain dormant unless the owner's secret key is applied to activate it. 
    This highlights a fundamental difference between \toolname and existing dynamic watermarking schemes: instead of directly verifying the trigger-set accuracy of the trained DNN $f(x;\theta)$,
    \textit{\toolname verifies the trigger-set accuracy of DNN $\hat{f} = f(x;\theta+\delta)$} where $\delta$ is the owner's \key and we call $f(x;\theta+\delta)$ the \unlocked.  In addition, $f$ is trained in such a way that it does not respond to the trigger.

    % Note the \key and the watermark trigger is pre-desgined by the user. 

\subsection{User Registration}
    We propose to include a trusted authority (TA) which applies a registration function $R$ to any user who wishes to register under the \toolname system. The trusted authority plays a similar role as the certificate authority in public-key infrastructure and can double as a judge in ownership disputes. $R$ takes a user-defined string as input and outputs a \key which is issued to the user. 
    %unique and used in both latent watermark embedding and latent watermark verification.
    
    The reason of having a trusted authority is to prevent users from registering a self-designed \key.
    A non-centralized registration could be built if all the users agree to use the same public one-way function \cite{goldreichfoundations} as the registration function.
    The property of a one-way function would make it extremely costly to register a self-designed \key.
    
    % It should be one-to-one and non-invertible (or it's extremely expensive to find the invertible function). 
\subsection{Latent Watermark Embedding}
    The watermark is latent in the sense that the network does not respond to it unless the \key is applied. 
    %The latent watermark is defined as a pre-designed trigger pattern which remains dormant unless latent watermark verification. 
    Latent watermarks are embedded during training using a specially crafted loss function. 
    We use $L_1(\theta)$ to denote the loss for DNN's normal operation on a clean dataset $D$ (with a label function $C_L$). 
    To embed a latent watermark, we use a latent watermark dataset $D_T$, which is a subset of clean data injected with the watermark (with a label function $T_L$) and a pre-determined \key $\delta$. To separate the DNN's normal operation from latent watermark verification, we define the latent watermark loss $L_2$ as the sum of following three terms:
    \begin{eqnarray}\label{eq: Watermark Loss}
        L_2(\theta) = \sum_{x\in D_T} L(f(x;\theta), C_L(x)) + \lambda_1\sum_{x\in D}L(f(x;\theta+\delta), C_L(x))\nonumber + \lambda_2\sum_{x\in D_T} L(f(x;\theta+\delta), T_L(x))
    \end{eqnarray}
    where $\lambda_1$, $\lambda_2$ are hyperparameters that control the importance of each term. 
    The first term is used to force the DNN not to respond to latent watermarks without the \key, thereby avoiding leaking any watermark information.
    The last two terms are designed for latent watermark verification which we will explain in more detail in Section~\ref{sec: Latent Watermark Verification}.
    Finally, the loss function for the DNN training process is a weighted sum of $L_1$ and $L_2$. 
    
\subsection{Latent Watermark Verification}\label{sec: Latent Watermark Verification}
    
    To claim the ownership of a DNN $f$, one should have a registered \key $\delta$, a latent watermark set $D_T$ and a trigger label function $T_L$. 
    Formally, the verification process is to check if
    \begin{eqnarray}
        % Pr_{x\in D_T} [\argmax f(x; \theta) = \argmax f(x; \theta+\delta)]\leq \frac{1}{c} + \epsilon \\
        Pr_{x\in D} [\argmax f(x; \theta+\delta) \not= C_L(x)]\leq \epsilon_1 \quad
        Pr_{x\in D_T} [\argmax f(x; \theta+\delta) \not= T_L(x)]\leq \epsilon_2
    \end{eqnarray}
    where $\theta$ is the parameter of $f$ and $\epsilon_1, \epsilon_2$ are user-defined thresholds. 
    
    For a latent watermarked DNN $f$, minimizing the last two terms in loss function~\ref{eq: Watermark Loss} force the perturbed DNN $\hat{f}$ to map any clean data $x$ to its label $C_L(x)$ and any latent watermark data to its trigger label $T_L(x)$. Therefore, the ownership of a DNN trained with loss function~\ref{eq: Watermark Loss} can be verified the owner with the corresponding \key.

\section{Analysis of \toolname}\label{sec: analysis}
\subsection{Performance Guarantee}
    In this section, we provide theoretical support on how \toolname's scheme of decoupling the DNN's normal operation and the backdoor-based ownership verification can preserve the performance of the target DNN.
    First, we show that any continuous function can be approximated by $f(x; \theta)$ where $f(x; \theta+\delta)$ is a backdoored DNN. 
    
    To this end, we consider a stronger version which requires showing that any $C(\mathbf{R}^{m+k}, \mathbf{R}^n)$ function can be approximated by $f(x; \theta+\delta)$, where we take $(x, \delta)\in \mathbf{R}^{m+k}$ as variables and $\theta$ as the parameter of this function. 
    
    \begin{theorem}[Universal Approximation Theorem for Parameter Space]
        Suppose $f(x; \theta)$ is a fully connected feedforward neural network with more than four hidden layers. $\delta$ is a perturbation for the weights of a single layer which is between the 2nd hidden layer and the last 2nd layer inclusive. We claim that the resulting neural network $f(x; \theta+\delta)$ is a universal approximation for any $C(\mathbf{R}^{m+k},\mathbf{R}^n)$, where $k$ is the dimension of $\delta$. That is, for any $g: \mathbf{R}^{m+k} \to \mathbf{R}^n$, there exists a $\theta$ such that $\sup ||f(x; \theta+\delta) - g(x, \delta)|| < \epsilon$.
    \end{theorem}
    
    \begin{corollary}[Performance Guarantee]\label{cor: Performance Guarantee}
        Suppose $f(x; \theta)$ is a feed-forward neural network with more than four hidden layers. For any non-zero weight perturbation $\delta$ which has the same size as the weight of a single layer between the 2nd hidden layer and the last 2nd layer inclusive. Then by the Universal Approximation Theorem for Parameter Space, there exists a \key $\delta$ such that $f(x; \theta)$ can approximate any continuous function where $f(x; \theta+\delta)$ is a backdoored DNN.
    \end{corollary}
    Corollary~\ref{cor: Performance Guarantee} also ensures that \toolname can work for any \key issued by the trusted authority.
    
    % \li{concern on the exists part; add a sentence to highlight that this ensures \toolname to work for any \key issued by the trusted authority}

\subsection{Defense against Backdoor Detection}\label{sec: Defense against Backdoor Detection}
    We consider backdoor detection techniques based on reverse-engineering the trigger and its target class. 
    %is to recover the potential trigger and the target class for a backdoor DNN. 
    Neural Cleanse~\cite{wang2019neural} is a good representative of these techniques. 
    It assumes a white-box model where the attacker has access to the DNN parameters and uses a gradient-based optimization method to reverse-engineer the trigger and its target class. 
    Such backdoor detection methods can be used by an attacker to extract the watermark information for a backdoor-based watermarked DNN.
    
    For \toolname, without the \key $\delta$, the watermarked DNN does not respond to the latent watermark. Therefore, for attackers who do not have knowledge of $\delta$, such backdoor detection methods will not identify the latent watermarks. An experiment in Section~\ref{sec: experiment} shows that the latent watermarked DNNs are indeed closer to clean DNNs than to backdoored DNNs.
    %when inspected by Neural Cleanse.
    % Immunization to backdoor detection is one of the most important reasons that we separate the DNN normal operation and ownership verification. Such separation makes the watermarked DNN safe under most backdoor detection (or watermark detection).
\subsection{Immunity to Surrogate Model Attacks}
    Surrogate model attacks target both static and dynamic watermarking schemes, since the watermark information (i.e. DNN's weights for static watermarking and the trigger and its target class for dynamic watermarking) could be leaked by performing input-output queries on the target model. On the contrary, a latent watermarked DNN does not respond to the watermarks unless the \key is applied and thus the watermark information would not be leaked through input-output queries. 
    
    In the following theorem, we claim that \toolname is immune to surrogate model attacks due to the arbitrariness of the \unlocked, i.e. it is not possible to recover $f(x; \theta+\delta)$ given only input-output queries of $f(x; \theta)$.
    \begin{theorem}[Immunity to Surrogate Model Attacks]
        % latent watermark is immune to surrogate model attacks. 
        Suppose $f(x;\theta)$ is a latent watermarked DNN with non-zero \key $\delta$, $D$ is the clean dataset and $D_T$ is the trigger set. For any $\epsilon > 0$, there exists another DNN $f(x;\theta')$:
        
        \quad 1. $Pr_{x\in D} [\argmax f(x; \theta) \not=\argmax f(x; \theta')]\leq \epsilon$;
        
        \quad 2. $Pr_{x\in D_T} [\argmax f(x; \theta+\delta) = \argmax f(x; \theta'+\delta)]\leq \frac{1}{c} + \epsilon$.
      
    \end{theorem}
    
\subsection{Immunity to Free-Rider Attacks}

    To the best of our knowledge, existing DNN OV techniques do not offer protection against free-rider attacks. Specifically, a (almost) free-rider can fine-tune (which is much less expensive than training) a pre-trained DNN which does not belong to them to embed their own watermarks and then claim ownership of the resulting DNN. 
    
    For \toolname, while we do not directly prohibit embedding watermarks through fine-tuning, the following theorem shows that using fine-tuning to embed a \textit{latent} watermark is as expensive as (re)training from scratch. Thus, there is no incentive for an attacker to forge their own latent watermarks on a pre-trained model this way.
    
    \begin{theorem}[Immunity to Free-Rider Attacks]
        If the \key $\delta$ is non-sparse and $||\delta||_2$ is large enough, then $\theta + \delta$ is close to a normal distribution on the whole parameter space, where $\theta$ is the parameter of a pre-trained DNN $f(x, \theta)$. Therefore, embedding $\delta$ into $f(x, \theta)$ via fine-tuning is as difficult as retraining from scratch.
    \end{theorem}
    % Note that fine-tuning to embed the latent watermark needs the access to both training data and trigger data while fine-tuning to embed the backdoor-based watermark only need the trigger data.
\section{Experiments}\label{sec: experiment}
In this section, we implement \toolname on a set of DNN models and evaluate their performances under backdoor detection, key guessing attacks, and watermark removal via model pruning and fine-tuning. The experiments are designed to answer the following questions: 

    1. How do the \toolname-watermarked DNNs perform relatively to normally trained DNNs on clean data? 
    %And will \toolname be harmful to the DNN's normal operation?
    
    2. How well does \toolname protect the watermarked models against key guessing attacks?
    
    3. Can the \toolname-watermarked DNNs evade backdoor detection methods such as Neural Cleanse?
    
    4. How well does the \toolname-watermarked DNNs resist watermark removal techniques such as model pruning and fine-tuning? 
    % \item How difficult to embed the AWP-based watermark via fine-tuning compare to backdoor-based watermark?

We use the following \textbf{evaluation metrics}:

1. \textbf{AC}: the accuracy of a watermarked DNN on clean test data; 

2. \textbf{AW}: the accuracy of a watermarked DNN on latent watermark data (the trigger set); 

3. \textbf{ACU}: the accuracy of a unlocked DNN (after the \key is applied) on clean test data; 

4. \textbf{AWU}: the accuracy of a unlocked DNN on latent watermark data.

\subsection{\toolname on MNIST/CIFAR10/GTSRB}
We train a \toolname-watermarked DNN on  MNIST~\cite{lecun1998mnist}/CIFAR10~\cite{krizhevsky2009learning}/GTSRB~\cite{Stallkamp2012} separately. For MNIST, we use a three-layer multilayer perceptron; for CIFAR10, we use vgg16\_comp~\cite{simonyan2014very}; for GTSRB, we use a six-layer convolutional neural network~\cite{o2015introduction}. More details about the DNN structure and the training hyperparameters can be found in the Supplemental Material.
We then verify each \toolname-watermarked DNN on both clean data and latent watermark data.

The results are presented in Table~\ref{tab: performance}. 
The performance of the \toolname-watermarked DNNs on clean data is very close to (and sometimes even better than) the performance of non-watermarked DNNs. For the watermark data, the unlocked DNNs $f(x, \theta+\delta)$ achieve >$99.93\%$ AWU for all cases.

%In short, \toolname preserves the performance of the models under normal operation while achieving a very high OV success rate.

%the performance of \toolname generated watermarked models are close to backdoor-based watermark meanwhile the \toolname generated watermark separates the DNN's normal operation and the OV.
\begin{table}
    \centering
    \resizebox{\columnwidth}{!}{%
    \begin{tabular}{c cccc|cccc|cccc}
    & \multicolumn{4}{c}{MNIST} & \multicolumn{4}{c}{CIFAR10} & \multicolumn{4}{c}{GTSRB}\\
    \#L & AC & AW & ACU & AWU & AC & AW & ACU & AWU & AC & AW & ACU & AWU \\
    \hline
    0& $98.03\%$ & $9.82\%$ & - & - &
    $78.67\%$ & $10.88\%$ & - & - &
    $91.79\%$ & $0.25\%$ & - & -\\
    1& $97.96\%$ & $9.85\%$ & $97.0\%$ & $\textbf{100.0\%}$ &
    $80.85\%$ & $10.48\%$ & $67.63\%$ & $99.99\%$ &
    $90.36\%$ & $0.43\%$ & $88.8\%$ & $99.97\%$\\
    2& $97.87\%$ & $9.8\%$ & $97.88\%$ & $\textbf{100.0\%}$ &
    $79.99\%$ & $8.95\%$ & $70.7\%$ & $\textbf{100.0\%}$ &
    $90.56\%$ & $0.63\%$ & $88.1\%$ & $99.93\%$\\
    3& $\textbf{98.07\%}$ & $9.83\%$ & $97.94\%$ & $\textbf{100.0\%}$ &
    $78.43\%$ & $10.06\%$ & $74.22\%$ & $99.99\%$ &
    $92.06\%$ & $0.4\%$ & $92.16\%$ & $99.98\%$\\
    4 & - & - & - & - &
    $\textbf{81.63\%}$ & $9.97\%$ & $80.51\%$ & $\textbf{100.0\%}$ &
    $\textbf{92.67\%}$ & $0.54\%$ & $92.46\%$ & $99.99\%$

    \end{tabular}
    %}
    }
    \caption{The performance of \toolname generated watermark on MNIST/CIFAR10/GTSRB. All models are train with 100 epochs. \#L: The layer that \key lies on (L=0 is the performance of a non-watermarked DNN).}
    \label{tab: performance}
\end{table}

\subsection{Defense against Key Guessing Attacks}\label{exp: ambiguity}
\begin{figure}
    \centering
    \includegraphics[width=0.9\textwidth]{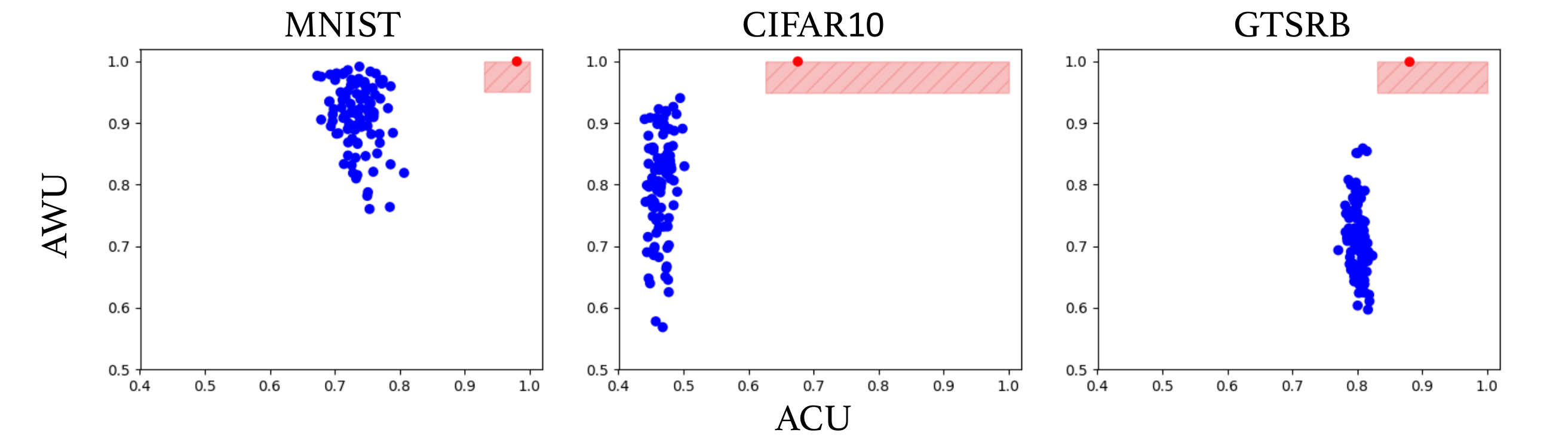}
    \caption{Defense against Key Guessing Attacks on MNIST/CIFAR10/GTSRB. The red dot corresponds to result of using the \key $\delta$. The blue dots correspond to results produced from using randomly guessed keys. We use $\epsilon_1 = \epsilon_2 = 0.05$ as the thresholds for latent watermark verification, i.e. any dot inside the red shaded area would be considered as verified.}
    \label{fig: ambiguity}
\end{figure}

We train a \toolname-watermarked DNN for each of the datasets MNIST/CIFAR10/GTSRB. We then sample 100 randomly generated weight key from a high dimensional uniform distribution (every entry is drawn from a uniform distribution between 0 and 1) and then perform ownership verification for the \toolname-watermarked DNN unlocked using a random weight key. 
From Figure~\ref{fig: ambiguity}, we can observe that all the randomly sampled keys fail the ownership verification.
%(i.e. none of the blue dots lies inside the red shaded area).

\subsection{Defense against Backdoor Detection}
We use the implementation of Neural Cleanse~\cite{wang2019neural} in TrojanZoo~\cite{pang:2022:eurosp} to evaluate how well \toolname-watermarked DNNs can evade backdoor detection. 
The GTSRB dataset is omitted because it is not supported by TrojanZoo. 
%We train a \toolname-watermarked DNN on MNIST/CIFAR10 (the code we use is from TrojanZoo \cite{pang:2022:eurosp} and it does not support GTSRB) and then apply Neural Cleanse \cite{wang2019neural} to \toolname-watermarked DNN/Clean DNN (a DNN trained by only clean data)/Dynamic Watermark~\cite{adi2018turning}. Neural Cleanse \cite{wang2019neural} is to recover the trigger using a gradient descend method. 
% We set the recover epochs=50 and learning rate=0.01. 
Table~\ref{tab: neural cleanse} shows that the detection results of Neural Cleanse.
We can observe that Neural Cleanse is not able to find any trigger with a small $L_1$ norm\footnote{Note that empirically speaking, for any DNN, there almost always exists a universal adversarial perturbation \cite{moosavi2017universal} with a large enough $L_1$ norm.} for both \toolname and Clean DNN (DNNs trained normally on only clean data). 
Thus, \toolname-watermarked DNNs would be treated as Clean DNNs when inspected by backdoor detection methods like Neural Cleanse.

%We conclude the \toolname-watermarked DNN is different from a dynamic watermarked DNN (backdoored DNN) and it's more close to a clean DNN.
\begin{table}
    \centering
    \scalebox{0.9}{
    \begin{tabular}{c| c ccc}
    % Dataset& \multicolumn{4}{c}{MNIST} & \multicolumn{4}{c}{CIFAR10} & \multicolumn{4}{c}{GTSRB}\\
    &  & \toolname & Clean DNN & BadNet~\cite{adi2018turning} \\
    \hline
    \multirow{3}{*}{MNIST} & ACC & 99.920\% &99.910\% & \textbf{100.000\%}\\ 
    & Norm & 47.635 & 77.725 & \textbf{2.177}\\
    & NormMD & 1.868 & 0.034 & \textbf{-3.438}\\
    \hline
    \multirow{3}{*}{CIFAR10} & ACC & 0.080\% & 0.010\% & \textbf{99.710\%} \\ 
    & Norm & 31.854 & 71.780 & \textbf{7.490}\\
    & NormMD & 0.186 & 1.924 & \textbf{-1.357}\\
    \\
    \end{tabular}
    %}
    }
    \caption{Neural Cleanse \cite{wang2019neural} results for \toolname/Clean DNN/BadNet on MNIST/CIFAR10. Clean DNN is a DNN trained using only clean data; BadNet~\cite{adi2018turning} is used as a backdoor-based dynamic watermarking scheme. ACC: backdoor accuracy on the reverse-engineered trigger. Norm: $L_1$ norm of the reverse-engineered trigger. NormMD: norm deviation of the reverse-engineered trigger (negative NormMD means the norm is less than the median and positive NormMD means the norm is greater than the median). A DNN is considered as a backdoored DNN if the backdoor accuracy is higher than some accuracy threshold (usually set to 90\%) and the NormMD is less than some NormMD threshold (usually set to -1).}
    \label{tab: neural cleanse}
\end{table}

\subsection{Resistance to Model Pruning}\label{sec: prune}
    We train a \toolname-watermarked DNN on MNIST/CIFAR10 and then apply model pruning \cite{cheng2017survey} to the \toolname-watermarked DNN. Pruning is applied to the last convolutional layer. We again use TrojanZoo for this evaluation~\cite{pang:2022:eurosp} and omit GTSRB since it is not yet supported by the tool. We report how the performances on clean data and watermarked data change as the ratio of pruning increases. Results for MNIST are included in Supplemental Material.
    
    In Figure~\ref{fig: prune}, we can observe that \toolname's performance is similar to that of clean DNN as the network is pruned -- almost no accuracy (AC) drop even when 95\% of neurons are pruned. 
    %(2) accuracy starts to drop only when 98\% neurons are pruned. 
    On the contrary, for backdoor-based watermarks~\cite{adi2018turning}, accuracy drops to 55\% when 95\% neurons are pruned. 
    In fact, the \toolname-watermarked DNNs can still pass ownership verification when 97\% neurons are pruned.
    
\begin{figure}
    \centering
    \includegraphics[width = 0.9 \textwidth]{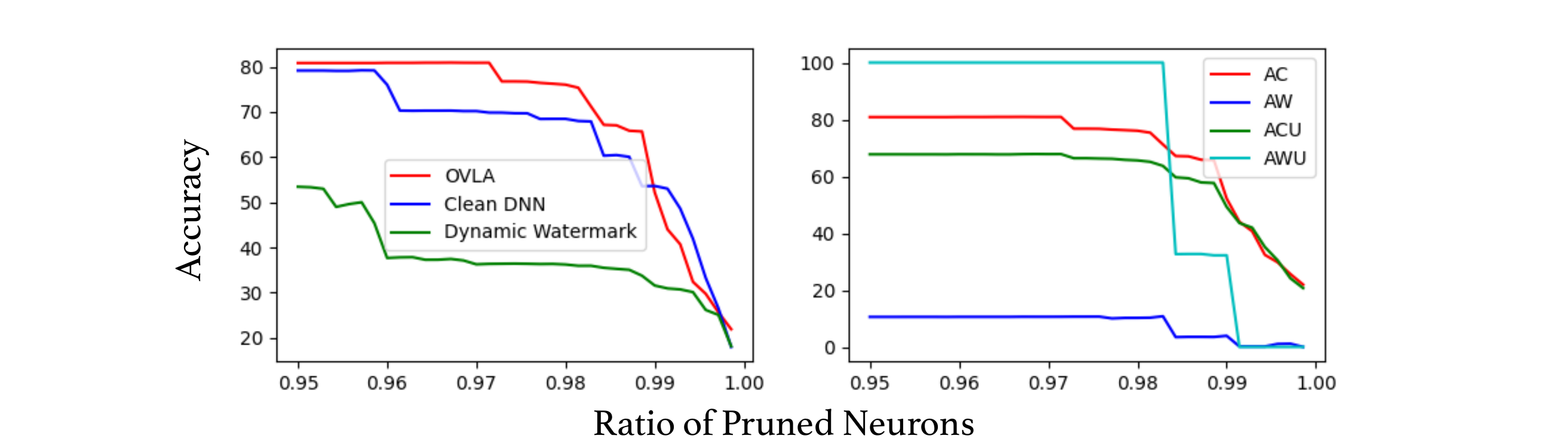}
    \caption{Left: Accuracy on clean data during model pruning for \toolname/Clean DNN/Dynamic Watermark~\cite{adi2018turning}. Right: Ownership verification during model pruning.}
    \label{fig: prune}
\end{figure}

\subsection{Resistance to Fine-Tuning}\label{sec: FT}
    \begin{table}[ht]
        \centering
        \scalebox{0.8}{
        \begin{tabular}{c cccc| cccc| cccc}
        & \multicolumn{4}{c}{Epoch 0} & \multicolumn{4}{c}{Epoch 50} & \multicolumn{4}{c}{Epoch 100}\\
        & AC & AW & ACU & AWU & AC & AW & ACU & AWU & AC & AW & ACU & AWU \\
        \hline
        MNIST & $97.96\%$ & $9.85\%$ & $97.0\%$ & $100.0\%$ & $98.02\%$ & $11.39\%$ & $95.38\%$ & $100.0\%$ & $98.0\%$ & $11.38\%$ & $92.18\%$ & $100.0\%$\\
        CIFAR10 & $80.85\%$ & $10.48\%$ & $67.63\%$ & $99.99\%$ &
        $80.44\%$ & $10.38\%$ & $66.26\%$ & $100.0\%$ &
        $80.14\%$ & $9.94\%$ & $66.11\%$ & $99.99\%$\\
        GTSRB & $90.78\%$ & $0.44\%$ & $89.02\%$ & $99.94\%$ &
        $90.74\%$ & $0.43\%$ & $89.11\%$ & $99.94\%$ &
        $90.74\%$ & $0.41\%$ & $89.11\%$ & $99.94\%$\\
        \end{tabular}
        %}
        }
        \caption{Ownership verification during fine-tuning on MNIST/CIFAR10/GTSRB.}
        \label{tab: FT}
    \end{table}
    We train a \toolname-watermarked DNN on MNIST/MNIST/CIFAR10 and then fine-tune the trained DNN with $20\%$ clean data from the training dataset.
    %\li{this setting may raise some questions; fine-tune only one layer or the whole network?}
    % \feisi{last layer or last convolution layer}.
    We then perform ownership verification on the fine-tuned DNN and record how its performance changes during fine-tuning. Results for using \key on the first layer
    %\li{our theorem guarantees between 2nd and 2nd-last layer though} 
    are reported in Table~\ref{tab: FT}. More result can be found in the Supplemental Material.
    Same as Section~\ref{exp: ambiguity}, we set $\epsilon = 5\%$ as the threshold of watermark verification. Only one fine-tuned DNN (at the 91st epoch for MNIST) failed the watermark verification.
        
% \subsection{Fine-Tuning to Embed Watermark}

\section{Conclusion}
In this paper, we propose \toolname, a novel DNN ownership verification methodology that decouples ownership verification from the DNN's normal operation. 
Our methodology has strong theoretical guarantees on normal operation performance, and defense against surrogate model attacks and free-rider attacks.
In addition, \toolname-watermarked DNNs show strong empirical resistance to backdoor detection, backdoor removal, key guessing-based ambiguity attacks. 
%Future works include further improving \toolname's resilience to key guessing attacks by including an extra loss term that penalizes keys dissimilar to the owner's \key, and a more extensive experimental evaluation.

\bibliographystyle{unsrtnat}
\bibliography{ref}  %%% Uncomment this line and comment out the ``thebibliography'' section below to use the external .bib file (using bibtex) .

\begin{thebibliography}{43}
\providecommand{\natexlab}[1]{#1}
\providecommand{\url}[1]{\texttt{#1}}
\expandafter\ifx\csname urlstyle\endcsname\relax
  \providecommand{\doi}[1]{doi: #1}\else
  \providecommand{\doi}{doi: \begingroup \urlstyle{rm}\Url}\fi

\bibitem[Krizhevsky et~al.(2012)Krizhevsky, Sutskever, and
  Hinton]{krizhevsky2012imagenet}
Alex Krizhevsky, Ilya Sutskever, and Geoffrey~E Hinton.
\newblock Imagenet classification with deep convolutional neural networks.
\newblock \emph{Advances in neural information processing systems},
  25:\penalty0 1097--1105, 2012.

\bibitem[Jo et~al.(2015)Jo, Hou, Eickholt, and Cheng]{jo2015improving}
Taeho Jo, Jie Hou, Jesse Eickholt, and Jianlin Cheng.
\newblock Improving protein fold recognition by deep learning networks.
\newblock \emph{Scientific reports}, 5\penalty0 (1):\penalty0 1--11, 2015.

\bibitem[Simonyan and Zisserman(2014)]{simonyan2014very}
Karen Simonyan and Andrew Zisserman.
\newblock Very deep convolutional networks for large-scale image recognition.
\newblock \emph{arXiv preprint arXiv:1409.1556}, 2014.

\bibitem[Floridi and Chiriatti(2020)]{floridi2020gpt}
Luciano Floridi and Massimo Chiriatti.
\newblock Gpt-3: Its nature, scope, limits, and consequences.
\newblock \emph{Minds and Machines}, 30\penalty0 (4):\penalty0 681--694, 2020.

\bibitem[Patterson et~al.(2021)Patterson, Gonzalez, Le, Liang, Munguia,
  Rothchild, So, Texier, and Dean]{patterson2021carbon}
David Patterson, Joseph Gonzalez, Quoc Le, Chen Liang, Lluis-Miquel Munguia,
  Daniel Rothchild, David So, Maud Texier, and Jeff Dean.
\newblock Carbon emissions and large neural network training.
\newblock \emph{arXiv preprint arXiv:2104.10350}, 2021.

\bibitem[Sada(2021)]{sada2021improving}
Samuel~O Sada.
\newblock Improving the predictive accuracy of artificial neural network (ann)
  approach in a mild steel turning operation.
\newblock \emph{The International Journal of Advanced Manufacturing
  Technology}, 112\penalty0 (9):\penalty0 2389--2398, 2021.

\bibitem[Uchida et~al.(2017)Uchida, Nagai, Sakazawa, and
  Satoh]{uchida2017embedding}
Yusuke Uchida, Yuki Nagai, Shigeyuki Sakazawa, and Shin'ichi Satoh.
\newblock Embedding watermarks into deep neural networks.
\newblock In \emph{Proceedings of the 2017 ACM on International Conference on
  Multimedia Retrieval}, pages 269--277, 2017.

\bibitem[Nagai et~al.(2018)Nagai, Uchida, Sakazawa, and
  Satoh]{nagai2018digital}
Yuki Nagai, Yusuke Uchida, Shigeyuki Sakazawa, and Shin’ichi Satoh.
\newblock Digital watermarking for deep neural networks.
\newblock \emph{International Journal of Multimedia Information Retrieval},
  7\penalty0 (1):\penalty0 3--16, 2018.

\bibitem[Adi et~al.(2018)Adi, Baum, Cisse, Pinkas, and Keshet]{adi2018turning}
Yossi Adi, Carsten Baum, Moustapha Cisse, Benny Pinkas, and Joseph Keshet.
\newblock Turning your weakness into a strength: Watermarking deep neural
  networks by backdooring.
\newblock In \emph{27th USENIX Security Symposium (USENIX Security 18)}, pages
  1615--1631, 2018.

\bibitem[Zhang et~al.(2018)Zhang, Gu, Jang, Wu, Stoecklin, Huang, and
  Molloy]{zhang2018protecting}
Jialong Zhang, Zhongshu Gu, Jiyong Jang, Hui Wu, Marc~Ph Stoecklin, Heqing
  Huang, and Ian Molloy.
\newblock Protecting intellectual property of deep neural networks with
  watermarking.
\newblock In \emph{Proceedings of the 2018 on Asia Conference on Computer and
  Communications Security}, pages 159--172, 2018.

\bibitem[Ma et~al.(2021)Ma, Chen, Hu, You, Xie, and Wang]{ma2021undistillable}
Haoyu Ma, Tianlong Chen, Ting-Kuei Hu, Chenyu You, Xiaohui Xie, and Zhangyang
  Wang.
\newblock Undistillable: Making a nasty teacher that cannot teach students.
\newblock \emph{arXiv preprint arXiv:2105.07381}, 2021.

\bibitem[Wang et~al.(2021)Wang, Xu, Xu, Wang, and Zhu]{wang2021non}
Lixu Wang, Shichao Xu, Ruiqi Xu, Xiao Wang, and Qi~Zhu.
\newblock Non-transferable learning: A new approach for model ownership
  verification and applicability authorization.
\newblock In \emph{International Conference on Learning Representations}, 2021.

\bibitem[Chen et~al.(2019)Chen, Rouhani, Fu, Zhao, and
  Koushanfar]{chen2019deepmarks}
Huili Chen, Bita~Darvish Rouhani, Cheng Fu, Jishen Zhao, and Farinaz
  Koushanfar.
\newblock Deepmarks: A secure fingerprinting framework for digital rights
  management of deep learning models.
\newblock In \emph{Proceedings of the 2019 on International Conference on
  Multimedia Retrieval}, pages 105--113, 2019.

\bibitem[Cheng et~al.(2017)Cheng, Wang, Zhou, and Zhang]{cheng2017survey}
Yu~Cheng, Duo Wang, Pan Zhou, and Tao Zhang.
\newblock A survey of model compression and acceleration for deep neural
  networks.
\newblock \emph{arXiv preprint arXiv:1710.09282}, 2017.

\bibitem[Wang and Kerschbaum(2019{\natexlab{a}})]{wang2019attacks}
Tianhao Wang and Florian Kerschbaum.
\newblock Attacks on digital watermarks for deep neural networks.
\newblock In \emph{ICASSP 2019-2019 IEEE International Conference on Acoustics,
  Speech and Signal Processing (ICASSP)}, pages 2622--2626. IEEE,
  2019{\natexlab{a}}.

\bibitem[Wang and Kerschbaum(2019{\natexlab{b}})]{wang2019robust}
Tianhao Wang and Florian Kerschbaum.
\newblock Robust and undetectable white-box watermarks for deep neural
  networks.
\newblock \emph{arXiv preprint arXiv:1910.14268}, 1\penalty0 (2),
  2019{\natexlab{b}}.

\bibitem[Chen et~al.(2021{\natexlab{a}})Chen, Chen, Zhang, and
  Wang]{chen2021you}
Xuxi Chen, Tianlong Chen, Zhenyu Zhang, and Zhangyang Wang.
\newblock You are caught stealing my winning lottery ticket! making a lottery
  ticket claim its ownership.
\newblock \emph{Advances in Neural Information Processing Systems}, 34,
  2021{\natexlab{a}}.

\bibitem[Yang et~al.(2021)Yang, Lao, and Li]{yang2021robust}
Peng Yang, Yingjie Lao, and Ping Li.
\newblock Robust watermarking for deep neural networks via bi-level
  optimization.
\newblock In \emph{Proceedings of the IEEE/CVF International Conference on
  Computer Vision}, pages 14841--14850, 2021.

\bibitem[Gu et~al.(2017)Gu, Dolan-Gavitt, and Garg]{gu2017badnets}
Tianyu Gu, Brendan Dolan-Gavitt, and Siddharth Garg.
\newblock Badnets: Identifying vulnerabilities in the machine learning model
  supply chain.
\newblock \emph{arXiv preprint arXiv:1708.06733}, 2017.

\bibitem[Le~Merrer et~al.(2020)Le~Merrer, Perez, and
  Tr{\'e}dan]{le2020adversarial}
Erwan Le~Merrer, Patrick Perez, and Gilles Tr{\'e}dan.
\newblock Adversarial frontier stitching for remote neural network
  watermarking.
\newblock \emph{Neural Computing and Applications}, 32\penalty0 (13):\penalty0
  9233--9244, 2020.

\bibitem[Fan et~al.(2019)Fan, Ng, and Chan]{fan2019rethinking}
Lixin Fan, Kam~Woh Ng, and Chee~Seng Chan.
\newblock Rethinking deep neural network ownership verification: Embedding
  passports to defeat ambiguity attacks.
\newblock \emph{Advances in Neural Information Processing Systems}, 32, 2019.

\bibitem[Boenisch(2020)]{boenisch2020survey}
Franziska Boenisch.
\newblock A survey on model watermarking neural networks.
\newblock \emph{arXiv preprint arXiv:2009.12153}, 2020.

\bibitem[Dong et~al.(2021)Dong, Yang, Deng, Pang, Xiao, Su, and
  Zhu]{dong2021black}
Yinpeng Dong, Xiao Yang, Zhijie Deng, Tianyu Pang, Zihao Xiao, Hang Su, and Jun
  Zhu.
\newblock Black-box detection of backdoor attacks with limited information and
  data.
\newblock In \emph{Proceedings of the IEEE/CVF International Conference on
  Computer Vision}, pages 16482--16491, 2021.

\bibitem[Wang et~al.(2019)Wang, Yao, Shan, Li, Viswanath, Zheng, and
  Zhao]{wang2019neural}
Bolun Wang, Yuanshun Yao, Shawn Shan, Huiying Li, Bimal Viswanath, Haitao
  Zheng, and Ben~Y Zhao.
\newblock Neural cleanse: Identifying and mitigating backdoor attacks in neural
  networks.
\newblock In \emph{2019 IEEE Symposium on Security and Privacy (SP)}, pages
  707--723. IEEE, 2019.

\bibitem[Craver et~al.(1998)Craver, Memon, Yeo, and Yeung]{craver1998resolving}
Scott Craver, Nasir Memon, B-L Yeo, and Minerva~M Yeung.
\newblock Resolving rightful ownerships with invisible watermarking techniques:
  Limitations, attacks, and implications.
\newblock \emph{IEEE Journal on Selected areas in Communications}, 16\penalty0
  (4):\penalty0 573--586, 1998.

\bibitem[Zhang et~al.(2020)Zhang, Chen, Liao, Zhang, Hua, and
  Yu]{zhang2020passport}
Jie Zhang, Dongdong Chen, Jing Liao, Weiming Zhang, Gang Hua, and Nenghai Yu.
\newblock Passport-aware normalization for deep model protection.
\newblock \emph{Advances in Neural Information Processing Systems},
  33:\penalty0 22619--22628, 2020.

\bibitem[Ong et~al.(2021)Ong, Chan, Ng, Fan, and Yang]{ong2021protecting}
Ding~Sheng Ong, Chee~Seng Chan, Kam~Woh Ng, Lixin Fan, and Qiang Yang.
\newblock Protecting intellectual property of generative adversarial networks
  from ambiguity attacks.
\newblock In \emph{Proceedings of the IEEE/CVF Conference on Computer Vision
  and Pattern Recognition}, pages 3630--3639, 2021.

\bibitem[Wu et~al.(2020)Wu, Xia, and Wang]{wu2020adversarial}
Dongxian Wu, Shu-Tao Xia, and Yisen Wang.
\newblock Adversarial weight perturbation helps robust generalization.
\newblock \emph{Advances in Neural Information Processing Systems},
  33:\penalty0 2958--2969, 2020.

\bibitem[Rakin et~al.(2020)Rakin, He, and Fan]{rakin2020tbt}
Adnan~Siraj Rakin, Zhezhi He, and Deliang Fan.
\newblock Tbt: Targeted neural network attack with bit trojan.
\newblock In \emph{Proceedings of the IEEE/CVF Conference on Computer Vision
  and Pattern Recognition}, pages 13198--13207, 2020.

\bibitem[Chen et~al.(2021{\natexlab{b}})Chen, Fu, Zhao, and
  Koushanfar]{chen2021proflip}
Huili Chen, Cheng Fu, Jishen Zhao, and Farinaz Koushanfar.
\newblock Proflip: Targeted trojan attack with progressive bit flips.
\newblock In \emph{Proceedings of the IEEE/CVF International Conference on
  Computer Vision}, pages 7718--7727, 2021{\natexlab{b}}.

\bibitem[Garg et~al.(2020)Garg, Kumar, Goel, and Liang]{garg2020can}
Siddhant Garg, Adarsh Kumar, Vibhor Goel, and Yingyu Liang.
\newblock Can adversarial weight perturbations inject neural backdoors.
\newblock In \emph{Proceedings of the 29th ACM International Conference on
  Information \& Knowledge Management}, pages 2029--2032, 2020.

\bibitem[Ning et~al.(2022)Ning, Li, Xin, Wu, and Wang]{ning2022hibernated}
Rui Ning, Jiang Li, Chunsheng Xin, Hongyi Wu, and Chonggang Wang.
\newblock Hibernated backdoor: A mutual information empowered backdoor attack
  to deep neural networks.
\newblock 2022.

\bibitem[Li et~al.(2021)Li, Wang, and Barni]{li2021survey}
Yue Li, Hongxia Wang, and Mauro Barni.
\newblock A survey of deep neural network watermarking techniques.
\newblock \emph{Neurocomputing}, 461:\penalty0 171--193, 2021.

\bibitem[Gou et~al.(2021)Gou, Yu, Maybank, and Tao]{gou2021knowledge}
Jianping Gou, Baosheng Yu, Stephen~J Maybank, and Dacheng Tao.
\newblock Knowledge distillation: A survey.
\newblock \emph{International Journal of Computer Vision}, 129\penalty0
  (6):\penalty0 1789--1819, 2021.

\bibitem[Lin et~al.(2019)Lin, Du, and Liu]{lin2019free}
Jierui Lin, Min Du, and Jian Liu.
\newblock Free-riders in federated learning: Attacks and defenses.
\newblock \emph{arXiv preprint arXiv:1911.12560}, 2019.

\bibitem[Goldreich()]{goldreichfoundations}
O~Goldreich.
\newblock Foundations of cryptography: Volume 1, basic tools, 2001.
\newblock \emph{Google Scholar Google Scholar Digital Library Digital Library}.

\bibitem[LeCun(1998)]{lecun1998mnist}
Yann LeCun.
\newblock The mnist database of handwritten digits.
\newblock \emph{http://yann. lecun. com/exdb/mnist/}, 1998.

\bibitem[Krizhevsky et~al.(2009)Krizhevsky, Hinton,
  et~al.]{krizhevsky2009learning}
Alex Krizhevsky, Geoffrey Hinton, et~al.
\newblock Learning multiple layers of features from tiny images.
\newblock 2009.

\bibitem[Stallkamp et~al.(2012)Stallkamp, Schlipsing, Salmen, and
  Igel]{Stallkamp2012}
J.~Stallkamp, M.~Schlipsing, J.~Salmen, and C.~Igel.
\newblock Man vs. computer: Benchmarking machine learning algorithms for
  traffic sign recognition.
\newblock \emph{Neural Networks}, pages~--, 2012.
\newblock ISSN 0893-6080.
\newblock \doi{10.1016/j.neunet.2012.02.016}.
\newblock URL
  \url{http://www.sciencedirect.com/science/article/pii/S0893608012000457}.

\bibitem[O'Shea and Nash(2015)]{o2015introduction}
Keiron O'Shea and Ryan Nash.
\newblock An introduction to convolutional neural networks.
\newblock \emph{arXiv preprint arXiv:1511.08458}, 2015.

\bibitem[Pang et~al.(2022)Pang, Zhang, Gao, Xi, Ji, Cheng, and
  Wang]{pang:2022:eurosp}
Ren Pang, Zheng Zhang, Xiangshan Gao, Zhaohan Xi, Shouling Ji, Peng Cheng, and
  Ting Wang.
\newblock Trojanzoo: Towards unified, holistic, and practical evaluation of
  neural backdoors.
\newblock In \emph{Proceedings of IEEE European Symposium on Security and
  Privacy (Euro S\&P)}, 2022.

\bibitem[Moosavi-Dezfooli et~al.(2017)Moosavi-Dezfooli, Fawzi, Fawzi, and
  Frossard]{moosavi2017universal}
Seyed-Mohsen Moosavi-Dezfooli, Alhussein Fawzi, Omar Fawzi, and Pascal
  Frossard.
\newblock Universal adversarial perturbations.
\newblock In \emph{Proceedings of the IEEE conference on computer vision and
  pattern recognition}, pages 1765--1773, 2017.

\bibitem[Hornik(1991)]{hornik1991approximation}
Kurt Hornik.
\newblock Approximation capabilities of multilayer feedforward networks.
\newblock \emph{Neural networks}, 4\penalty0 (2):\penalty0 251--257, 1991.

\end{thebibliography}

\section{Supplemental Material}\label{sec: Supplemental Material}

\subsection{Broader Impacts}
Deep neural network models can be very expensive to train. 
They are valuable intellectual properties (IPs) to their owners and thus it is important to protect them from misuse such as illegal copying,  re-distribution, and free-riding. 
This paper presents a new methodology for neural network ownership verification with strong guarantees against a variety of attacks. 
It is worth noting that the goal of a methodology like is to protect IPs, and it does not hinder the sharing or distribution of neural network models. The broader debate about the issue of ownership of IPs is outside the scope of this paper.
In terms of the technique presented, given the connection between dynamic watermarking and neural backdoor attacks, it would be fair to ask whether the proposed technique could be exploited by an attacker. 
More specifically, an attacker could use \toolname to insert a latent Trojan during training, and then activate it by applying the secret weight key (using techniques such as bit-flip attacks) during deployment. 
While this work calls attention to this new model of neural backdoor attacks, this attack requires a strong attacker model -- the attacker needs to have access to and the ability to manipulate the network both during training and deployment, which is unlikely in most settings.

\subsection{Proofs of Theorems}
    \setcounter{theorem}{0}
    \setcounter{corollary}{0}
    \begin{theorem}[Universal Approximation Theorem for Parameter Space]\label{apx: Universal Approximation Theorem for Parameter Space}
        Suppose $f(x; \theta)$ is a fully connected feedforward neural network with more than four hidden layers. $\delta$ is a perturbation for the weights of a single layer which is between the 2nd hidden layer and the last 2nd layer inclusive. We claim that the resulting neural network $f(x; \theta+\delta)$ is a universal approximation for any $C(\mathbf{R}^{m+k},\mathbf{R}^n)$, where $k$ is the dimension of $\delta$. That is, for any $g: \mathbf{R}^{m+k} \to \mathbf{R}^n$, there exists a $\theta$ such that $\sup ||f(x; \theta+\delta) - g(x, \delta)|| < \epsilon$.
    \end{theorem}
    
    \begin{proof}
        For any $g: \mathbf{R}^{m+k} \to \mathbf{R}^n$ and $\epsilon > 0$, by Universal Approximation Theorem \cite{hornik1991approximation}, there exists a neural network $h^1(x, \delta; \theta_1)$, such that $||h^1(x, \delta; \theta_1) - g(x, \delta)|| < \frac{\epsilon}{2}$. 
        
        % Now we need to construct a neural network $f(x; \theta+\delta)$ which takes $x$ as input and $\theta$ as weight parameter and $||f(x; \theta+\delta) - h^1(x, \delta; \theta_1)||< \frac{\epsilon}{2}$. 
        
        We write the first linear layer of $ h^1(x, \delta; \theta_1)$ as $y = W_1 x + W_2 \delta + b$, where $W_1, W_2$ and $b$ are parameter of the first linear layer. By the Universal Approximation Theorem~\cite{hornik1991approximation}, we can construct a neural network $h^2(x; \theta_2)$ with an output dimension equal to $[x, W_2]$ (by concatenating $x$ and $W_2$) and $||h^2(x; \theta_2) - [x, W_2]|| < \frac{\epsilon}{2L}$, where $L$ is the Lipschitz constant of $h^1(x, \delta; \theta_1)$.
        
        Then we build up a neural network using $h^2(x; \theta_2)$, a linear layer $l(x, W_2) = W_1 x + W_2\delta + b$ and $h^1_{[1:]}(y; \theta_1)$, which is $h^1(x, \delta; \theta_1)$ removing the first linear layer. 
        \begin{eqnarray}
            f(x; \theta+\delta) = h^1_{[1:]}(y; \theta_1) \circ l \circ h^2(x; \theta)
        \end{eqnarray}
        We claim that this is a neural network, since both $h^1_{[1:]}(y; \theta_1)$ and $h^2(x; \theta)$ are neural networks and $l$ is a linear function. By the construction, we have $||f(x; \theta+\delta) - h^1_{[1:]}(y, \theta_1) \circ l(x, \delta)||< \frac{\epsilon}{2}$. Combining with $||h^1(x, \delta; \theta_1) - g(x, \delta)|| = ||h^1_{[1:]}(y, \theta_1) \circ l(x, \delta) - g(x, \delta)|| < \frac{\epsilon}{2}$, we have $||f(x; \theta+\delta) - g(x, \delta)||< \epsilon$.
    \end{proof}
    
    \begin{corollary}[Performance Guarantee]\label{apxcor: Performance Guarantee}
        Suppose $f(x; \theta)$ is a feed-forward neural network with more than four hidden layers. For any non-zero weight perturbation $\delta$ which has the same size as the weight of a single layer between the 2nd hidden layer and the last 2nd layer inclusive. Then by the Universal Approximation Theorem for Parameter Space, there exists a \key $\delta$ such that $f(x; \theta)$ can approximate any continuous function where $f(x; \theta+\delta)$ is a backdoored DNN.
    \end{corollary}
    
    \begin{proof}
        By the definition of continuous function, for any $\epsilon$, $\delta$ and a continuous function $h\in C(\mathbf{R}^{m}, \mathbf{R}^n)$, there exists a $C(\mathbf{R}^{m+k}, \mathbf{R}^n)$ function $g(x, \delta)$ such that $|g(x, 0) - h(x)|<\epsilon$ and $g(x, \delta)$ is a backdoored DNN (has high accuracy on trigger data). 
        
        By Theorem~\ref{apx: Universal Approximation Theorem for Parameter Space}, there is a neural network $f(x; \theta+\delta)$, such that $f(x; \theta+\delta)$ is an approximation of $g(x, \delta)$. Therefore, $f(x; \theta)$ could approximate any continuous function where $f(x; \theta + \delta)$ is a backdoored DNN. 
    \end{proof}

    \begin{theorem}[Immunity to Surrogate Model Attacks]
        Suppose $f(x;\theta)$ is a latent watermarked DNN with non-zero \key $\delta$, $D$ is the clean dataset and $D_T$ is the trigger set. For any $\epsilon > 0$, there exists another DNN $f(x;\theta')$:
        \begin{enumerate}
            \item $Pr_{x\in D} [\argmax f(x; \theta) \not=\argmax f(x; \theta')]\leq \epsilon$;
            \item $Pr_{x\in D_T} [\argmax f(x; \theta+\delta) = \argmax f(x; \theta'+\delta)]\leq \frac{1}{c} + \epsilon$.
        \end{enumerate}
    \end{theorem}
    
    \begin{proof}
        We fix $\theta$ and view $\delta$ as an input variable, then $f(x; \theta+\delta)$ is a function from $\mathbf{R}^{m+k}$ to $\mathbf{R}^n$, where $k$ is the dimension of $\delta$. 
        
        Since $f(x; \theta+\delta)\in C(\mathbf{R}^{m+k}, \mathbf{R}^n)$ is a continuous function, there must exist another continuous function $h(x, \delta)\in C(\mathbf{R}^{m+k}, \mathbf{R}^n)$. such that
        \begin{enumerate}
            \item $Pr_{x\in D} [\argmax f(x; \theta) \not=\argmax h(x, 0)]\leq \epsilon$ and for any input $x\in D$, $h_i(x, 0) > \max_{j\not=i} h_j(x, 0) + \epsilon$, where $i = \argmax h(x, 0)$;
            \item $Pr_{x\in D_T} [\argmax f(x; \theta+\delta) = \argmax h(x, \delta)]\leq \frac{1}{c} + \epsilon$ and for any input $x\in D_T$, $h_i(x, \delta) > \max_{j\not=i} h_j(x, \delta) + \epsilon$, where $i = \argmax h(x, \delta)$.
        \end{enumerate}
        By Theorem~\ref{apx: Universal Approximation Theorem for Parameter Space}, there is a neural network $f(x; \theta'+\delta)$, such that $|f(x; \theta'+\delta) - h(x, \delta)| < \frac{1}{2}\epsilon$, then we have 
        \begin{enumerate}
            \item for any input $x\in D$, $f_i(x; \theta') > h_i(x, 0)-\frac{1}{2}\epsilon > \max_{j\not=i} h_j(x, 0) + \frac{1}{2}\epsilon > \max_{j\not=i} f_j(x; \theta')$, where $i = \argmax h(x, 0)$. Therefore, $\argmax h(x, 0) = \argmax f(x; \theta')$ and $Pr_{x\in D} [\argmax f(x; \theta) \not=\argmax f(x; \theta')]\leq \epsilon$;
            \item for any input $x\in D_T$, $f_i(x; \theta'+\delta) > h_i(x, \delta)-\frac{1}{2}\epsilon > \max_{j\not=i} h_j(x, \delta) + \frac{1}{2}\epsilon > \max_{j\not=i} f_j(x; \theta'+\delta)$, where $i = \argmax h(x, \delta)$. Therefore, $\argmax h(x, \delta) = \argmax f(x; \theta'+\delta)$ and $Pr_{x\in D_T} [\argmax f(x; \theta+\delta) = \argmax f(x; \theta'+\delta)]\leq \frac{1}{c} + \epsilon$.
        \end{enumerate}
    \end{proof}

    \begin{theorem}[Immunity to Free-Rider Attacks]
        If the \key $\delta$ is non-sparse and $||\delta||_2$ is large enough, then $\theta + \delta$ is close to a normal distribution on the whole parameter space, where $\theta$ is the parameter of a pre-trained DNN $f(x, \theta)$. Therefore, embedding $\delta$ into $f(x, \theta)$ via fine-tuning is as difficult as retraining from scratch.
    \end{theorem}
    
    \begin{proof}
        Consider $f(x; \theta)$ as a neural network that achieves a local minimum for a certain loss function on the clean data. Since our \key $\delta$ is independent with $f(x; \theta)$, we can consider $\delta$ as a random vector with zero mean and $\sigma_\delta$ standard deviation.
        
        Assume that there is an attacker who wants to fine-tune $f(x; \theta)$ such that the resulting neural network $f(x; \hat{\theta})$ has the following properties:
        \begin{enumerate}
            \item both $f(x; \hat{\theta})$ and $f(x; \hat{\theta}+\delta)$ have high accuracy on the clean data;
            \item $f(x; \hat{\theta}+\delta)$ has high accuracy on the latent watermark data;
        \end{enumerate}
        For a DNN structure, there is a large number of local minima that can be achieved. $\theta$ can be viewed as a random vector that takes one of $\{\theta_i\}$ with some probability.
        By the Law of Large Numbers, with a large number of local minima, the distribution of $\theta+\delta$ is close to a normal distribution with $\mu = \sum_{i=1}^{i=n} p_i\theta_i$ and $\sigma = O(\sigma_\delta)$. Since $\sigma_\delta$ is large enough. We have $\theta+\delta$ is close to a normal distribution on the whole parameter space. 
        Therefore, training with initial weights $\theta+\delta$ is the same as training from scratch (i.e. from a randomly initialized $\theta'+\delta$).
    \end{proof}
        
    \subsection{Additional Experiment Details}\label{apxsec: Additional Experiment Details}
    
    % Our code is available at\\ \url{https://drive.google.com/drive/folders/1EUThz46GJLng4jSUhNgLnOC6nZWGO6cH?usp=sharing}.
    
    \textbf{Experiment Platform}
        
    All experiments were run on a ten-core 2.6 GHz Intel Xeon E5-2660v3 with 128 GB of memory and a NVIDIA Tesla K40m GPU.
    
    \textbf{DNNs Used in the Experiments:}
    
    The classifier for MNIST is a three-layer multilayer perceptron with 128 neurons in each hidden layer.
    
    The classifier for CIFAR10 is a vgg16\_comp model from ~\cite{simonyan2014very}.
    
    The classifier for GTSRB contains three hidden convolutional layers with 32/64/64 features respectively and three fully connected layers with 256 neurons in each layer. A max pooling layer is used after every hidden convolutional layer.
    
    \textbf{Dataset Used in the Experiments:}
    
    MNIST~\cite{lecun1998mnist} is a database of handwritten digits which has a training set of 60,000 examples and a test set of 10,000 examples.
    
    CIFAR10~\cite{krizhevsky2009learning} is a dataset consists of 50,000 training images and 10,000 test images. 
    
    The German Traffic Sign Recognition Benchmark (GTSRB~\cite{Stallkamp2012}) is a dataset of traffic signs which has 39,209 training images and 12,630 test images.
    
    \textbf{Hyperparameters Used in the Experiments:}
    
    The learning rate is set to $10^{-3}$ for both training \toolname-watermarked DNNs and training non-watermarked DNNs.
    
    We set $\lambda_1 = \lambda_2 = 1$ (in loss function~\ref{eq: Watermark Loss}) for training \toolname-watermarked DNNs.
    
    The \key and the latent watermark we used in the experiments is randomly generated from a high-dimensional uniform distribution (i.e. every entry is drawn from a uniform distribution between 0 and 1).
    
    The position of the latent watermark is at the up-left corner and the size is $6 \times 6$. 
    \begin{figure}[H]
        \centering
        \includegraphics[width = 0.9 \textwidth]{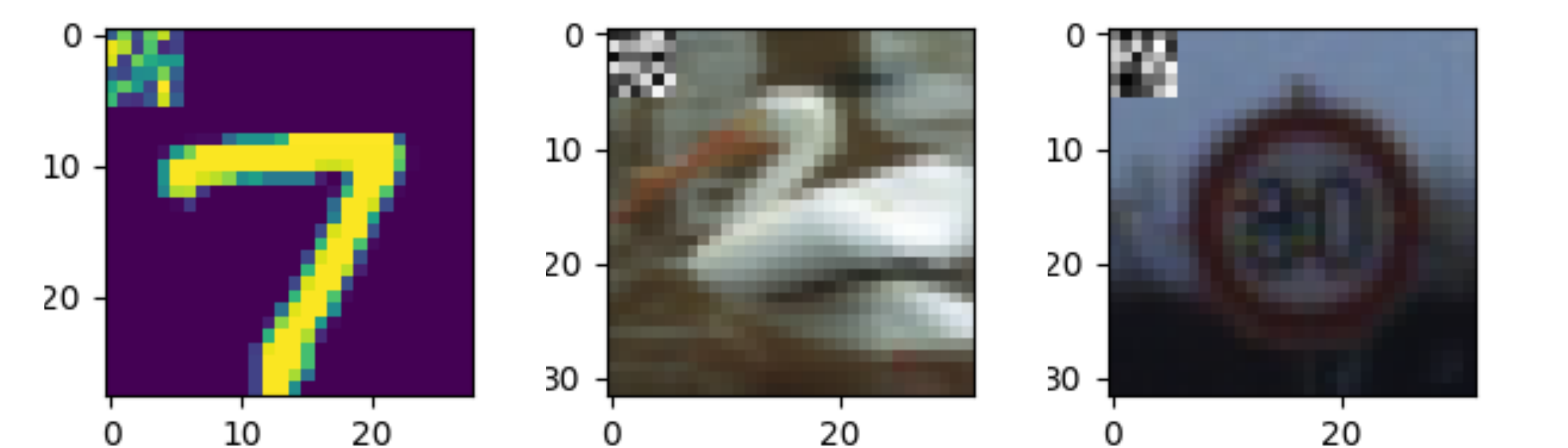}
        \caption{samples of image with latent watermark}
        \label{apxfig: sample}
    \end{figure}
        
    For Neural Cleanse~\cite{wang2019neural} used in Section~\ref{sec: Defense against Backdoor Detection}, the learning rate is set to $10^{-3}$ and the maximum epoch is set to 50. 
    
    For Fine-Tuning used in Section~\ref{sec: FT}, the learning rate is set to $2*10^{-4}$ and the maximum epoch is set to 100. 
    
    \textbf{Additional Details for Resistance to Model Pruning}
    
\begin{figure}[ht]
        \centering
        \includegraphics[width = 0.7 \textwidth]{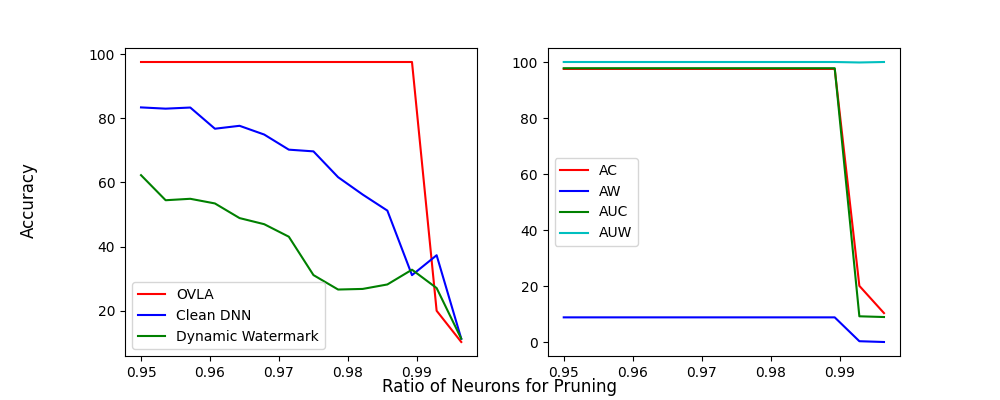}
        \caption{Left: accuracy on clean data during Neural Pruning on \toolname/Clean DNN/Dynamic Watermark. Right: Ownership verification during Neural Pruning.}
        \label{apxfig: prune}
    \end{figure}

    Figure~\ref{apxfig: prune} is the result for Section~\ref{sec: prune} on MNIST. Similar to the result shown in Figure~\ref{fig: prune}, we observe that \toolname has a strong resistance to Model Pruning.
    
    \textbf{Additional Details for Resistance to Fine-Tuning}
    
\begin{figure}[ht]
        \centering
        \includegraphics[width = 0.99 \textwidth]{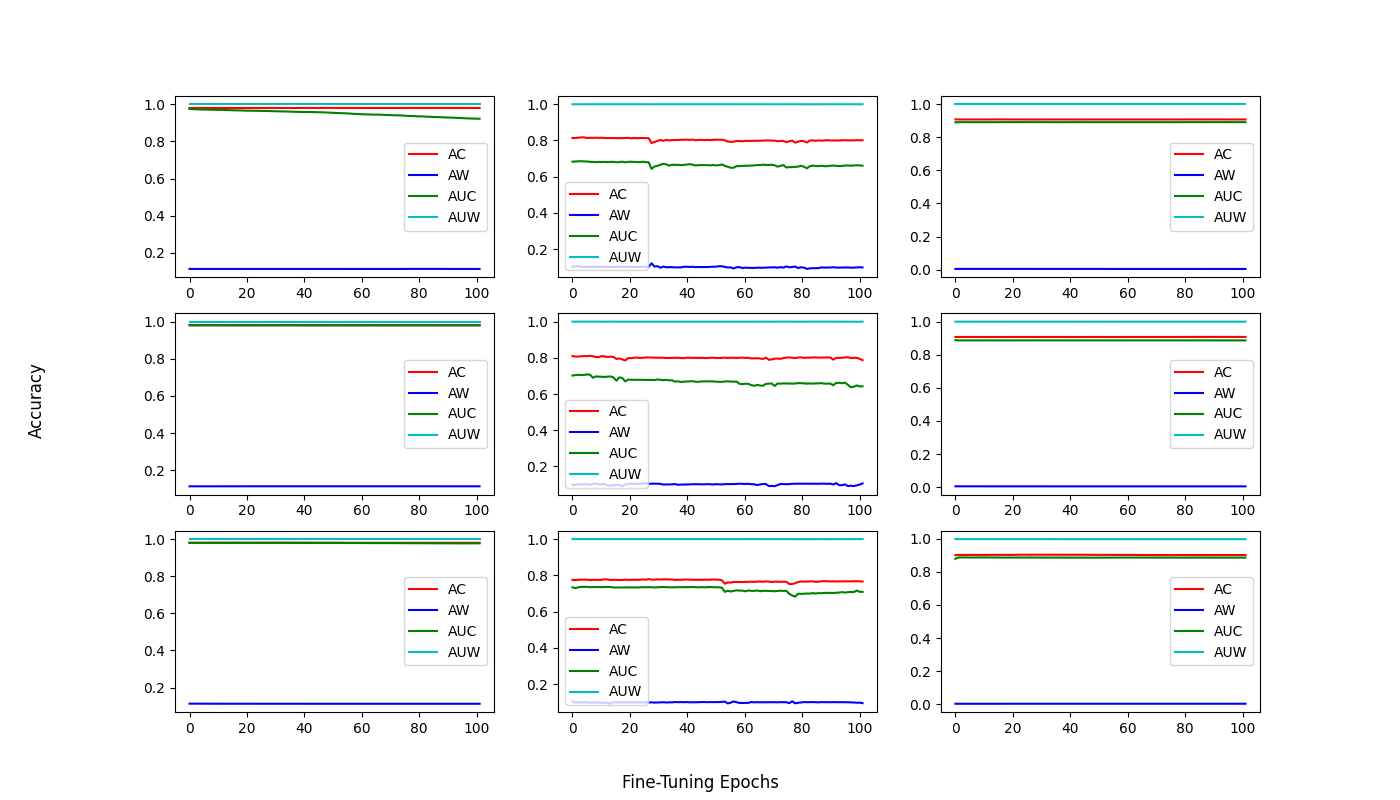}
        \caption{Ownership Verification during Fine-Tuning. Rows: MNIST/CIFAR10/GTSRB. Columns: \key on first/second/third layer.}
        \label{apxfig: FT}
    \end{figure}

    Figure~\ref{apxfig: FT} is the full results for Section~\ref{sec: FT}. After 100 epoch fine-tuning, all the metrics (AC, AW, ACU and AWU) drop slightly. We set $\epsilon = 5\%$ as the threshold of watermark verification. Among all nine \toolname-watermarked DNNs, only one fine-tuned DNN (at the 91st epoch for MNIST) failed the watermark verification.
    
\end{document}